\documentclass[amsmath,amssymb,showpacs,showkeywords,twocolumn]{revtex4}
\usepackage[dvips]{graphicx,color}
\usepackage{times}
\usepackage{mathrsfs}
\usepackage{amsmath}
\usepackage{graphicx}
\usepackage{epsfig}
\usepackage{dcolumn}
\usepackage{bm}
\usepackage{multirow}
\usepackage{xcolor}
\begin{document}
\title{Analytic modelling of Quantum Capacitance and Carrier Concentration for $\beta_{12}$ - Borophene FET based Gas Sensor}
\author{Nimisha Dutta\footnote{nimishadutta@dibru.ac.in}, Reeta Devi\footnote{reetadevi@dibru.ac.in}, Arindam Boruah\footnote{arindamboruah@dibru.ac.in} and Saumen Acharjee\footnote{saumenacharjee@dibru.ac.in}}
\affiliation{Department of Physics, Dibrugarh University, Dibrugarh 786 004, 
Assam, India,}

\begin{abstract}
In this work, we investigate the physical and electronic properties of $\beta_{12}$ - borophene FET-based gas sensor using a theoretical quantum capacitance model based on tight - binding approach. We study the impact of adsorbed NH$_3$, NO, NO$_2$ and CO gas molecule on its density of states, carrier concentration, quantum capacitance and I-V characteristics. We found a remarkable variation in the energy band structure and the density of states (DOS) of the $\beta_{12}$ - borophene in the presence of the adsorbed gas molecule. The appearance of non-identical Van-Hove singularities in the DOS in the presence of adsorbed gas molecules strongly indicates the high sensitivity of $\beta_{12}$ - borophene. We found a significant increase in the carrier concentration for NH$_3$ gas while it decreases for all other gases. Moreover, a drastic change in quantum capacitance and current-voltage relation is also observed in adsorbed gases. The different properties of the given gas molecules are compared with the pristine borophene and found to exhibit distinct wrinkles in each case, thereby indicating the strong selectivity of our proposed gas sensor. Though $\beta_{12}$ - borophene is found to be highly sensitive for all studied gases, the NO gas is found to be most sensitive compared to the others. 
\end{abstract}

\maketitle

\section{Introduction} 
During the last two decades, two-dimensional (2D)  material has garnered enormous interest due to their unique physical properties, making them a promising candidate for next-generation electronics and energy conversion devices \cite{novoselove,lemme,liu,xu,quellmalz,illarionov,passi,fuechsle,acharjee}. As the first 2D  material with significantly different electronic \cite{balandin,lee,zhang,sang}, thermal \cite{sang,pop}, mechanical \cite{papageorgiou} and optical properties \cite{papageorgiou}, graphene-based devices have been widely explored. However, the absence of band gap, low on/off ratio and extremely high carrier mobility in graphene restrict its use as a semiconducting system \cite{neto,mccann,ando}. So, there is a need for new alternative materials with similar but more advanced electronic properties than graphene. Thus a class of new 2D   materials, like phosphorene \cite{akhtar}, silicene \cite{vogt}, stanene \cite{zhu2}, antimonene \cite{ji}, bismuthene \cite{drozdov}, germanine \cite{davila}, molybdenum disulfide \cite{splendiani} etc., have been synthesized as alternate candidates for graphene in recent times. Depending on their stacking and composition, these 2D  materials can possess distinctive physicochemical properties, such as electrical conductivity, thermal conductivity, and band structure around the Fermi level, making them suitable for various applications \cite{akhtar,vogt,zhu2,ji,drozdov,davila,splendiani}.

Alongside the other promising candidates,  Borophene, a single element 2D sheet of boron, has also been synthesized on Ag (111) substrate in recent times \cite{mannix,zhang2,feng}. Unlike graphene, borophene has a triangular and hexagonal array of atoms exhibiting in-plane elasticity and can be more flexible in some configurations. Apart from that, borophene has high mobility, electrical and thermal conductivity and also possesses high mechanical strength, with Young's modulus much higher than its rivals \cite{zhang3,mozvashi,yuan}. Thus borophene has excellent potential in advanced electrical information, sensing and other optical, mechanical and thermal applications \cite{wang}. Moreover, previous works indicate that the borophene-based systems are active to hazardous gases, thus making them suitable for gas sensing applications \cite{huang,hou,shukla}. Significant attention also has been given to the optimization of the borophene/MoS$_2$ heterostructure using density functional theory (DFT) implemented in the VASP package \cite{shen}. Although there exist several boron clusters with fascinating properties, it was found that the allotrope $\beta_{12}$ - borophene is thermodynamically most stable as compared to the other members of the borophene family \cite{peng,feng}. It is observed that $\beta_{12}$ - borophene can display semiconducting and semi-metallic properties in the presence of an applied electric field and charged impurity \cite{li2}. Furthermore, this boron cluster can also display anisotropic Kubo conductivity \cite{nobahari} in the presence of an applied electric field, making them unique from other members. Though several attempts have been made to understand the adsorption properties in different boron-based clusters both theoretical as well as experimentally \cite{huang,hou,shukla,opoku}, the study of gas sensing in $\beta_{12}$ - borophene is limited. So, in this work, we have focused on $\beta_{12}$ phase of borophene to investigate the adsorption of the gas molecules. 

Recently, quantum capacitance has gained much attention in studying 2D electron systems to reveal interesting many-body effects \cite{tsu, xia, hassanzadazar}. It also carries information regarding the  ground state of the system, revealing the effect of electron-electron interaction and quantum correlation. The Quantum capacitance of a system carries essential information regarding the density of states. Recently quantum capacitance model has been used to measure the Van-Hove singularities and Luttinger parameters in a one-dimensional system \cite{li4,yu3}. Moreover, the quantum capacitance measurements have also shown the linear density of states of topological insulator Bi$_2$Se$_3$ \cite{inhofer} and monolayer graphene \cite{cheremisin, pourasl,dutta}. Thus, it is necessary to have knowledge about the quantum capacitance for understanding the system properly. Though quantum capacitance has been widely studied theoretically and experimentally in graphene, such studies have been limited to borophene until now. An attempt has been made to explore the quantum capacitance effect in graphene chemical sensors and $\delta_6$ - borophene \cite{kolavada,john} to induce changes in the capacitance values upon the adsorption of molecules on the surface. But quantum capacitance of $\beta_{12}$ - borophene has not been explored so far. Therefore, in this work, we investigated the quantum capacitance of $\beta_{12}$ - borophene in the presence of adsorbed gas molecules.

The organisation of the paper is as follows: we present a theoretical framework based on tight-binding (TB) Hamiltonian and also present a quantum capacitance model of $\beta_{12}$ - borophene-based FET based gas sensor in Section II. In Section III, we study the density of states, carrier concentration, quantum capacitance and I-V characteristics to understand the gas sensing property of $\beta_{12}$ - borophene-based FET. A summary of our work is presented in Section IV.

\section{Theoretical framework}
\begin{figure}[hbt]
\centerline
\centerline{
\includegraphics[scale = 0.6]{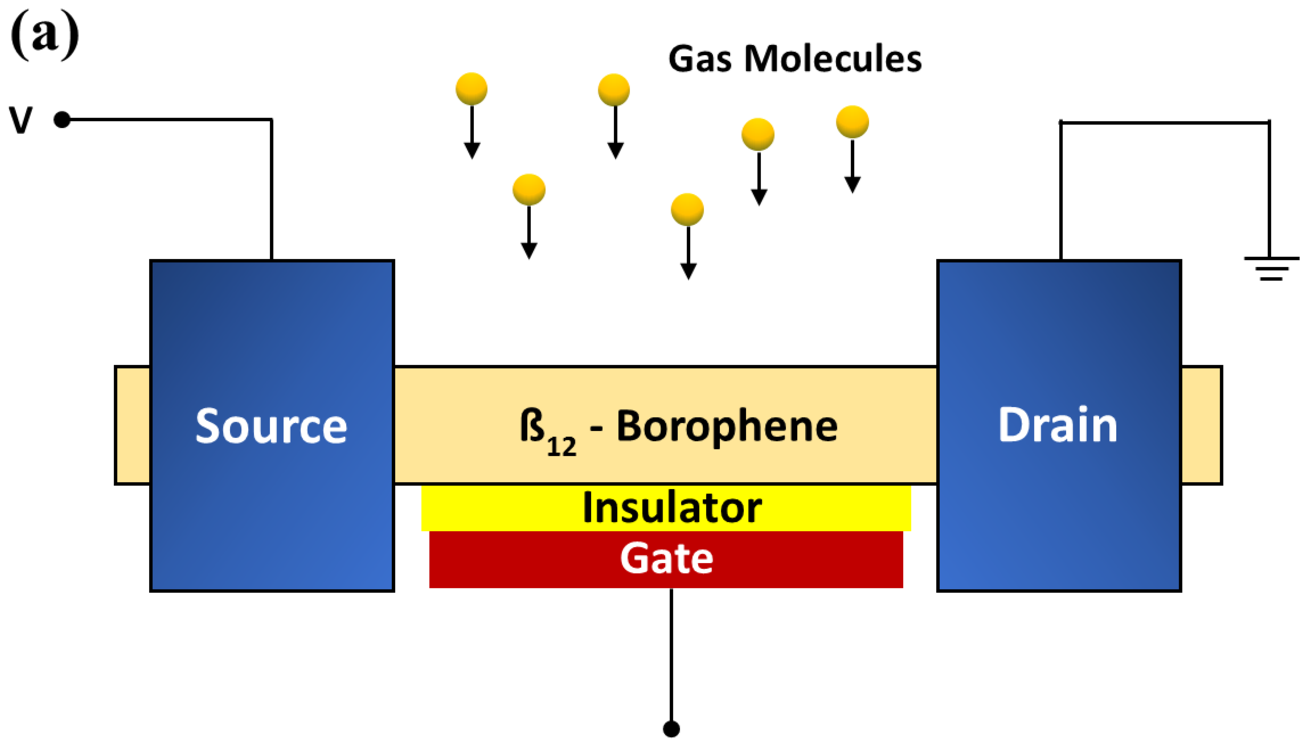}
\includegraphics[scale = 0.6]{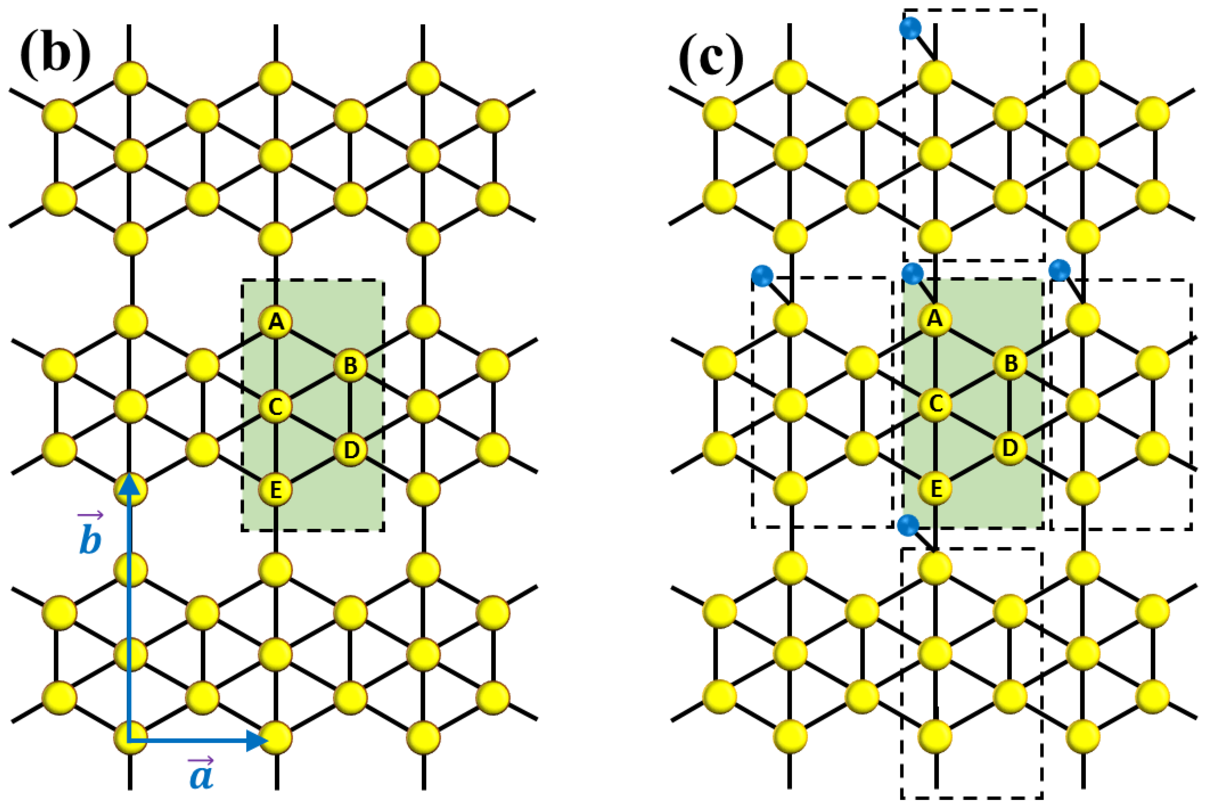}
\includegraphics[scale = 0.5]{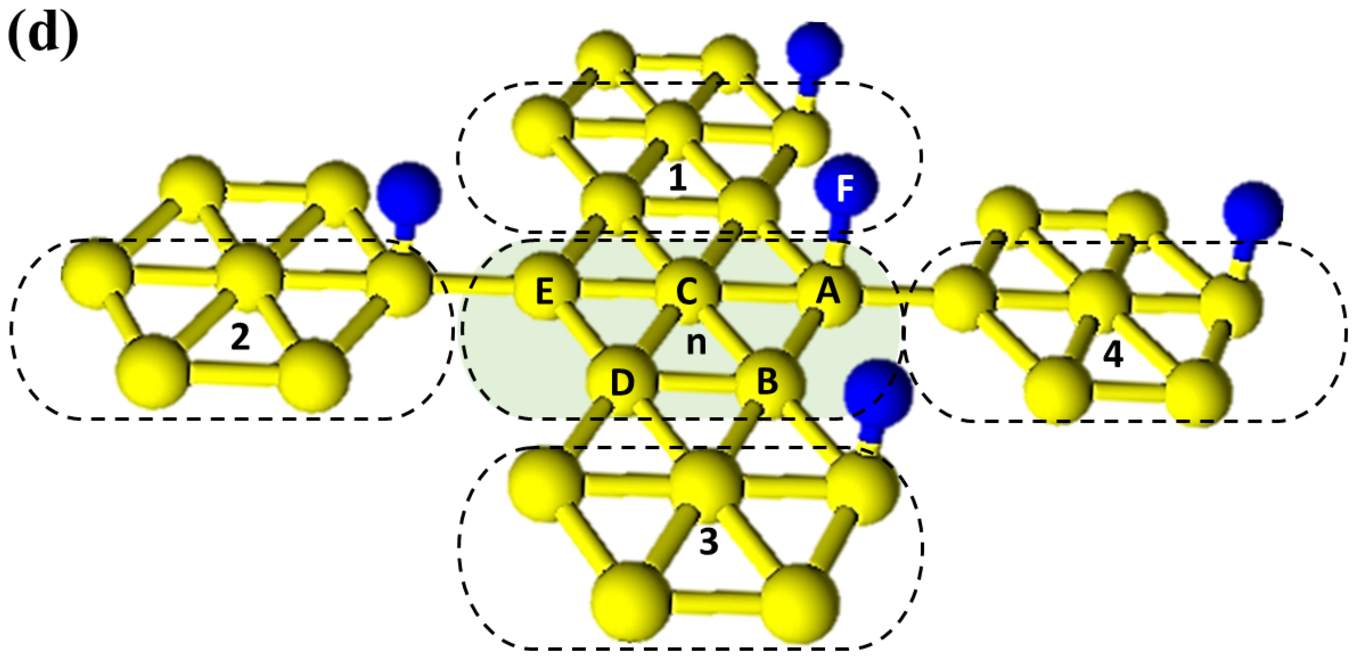}
\hspace{-0.09cm}
}
\caption{(a) Schematic illustration of $\beta_{12}$ - borophene field effect transistor (FET) based gas sensor. (b) Top view of the geometry structure of $\beta_{12}$ - borophene. The unit cell (shaded green region) is one-half of the honeycomb lattice and consists of five boron atoms. The lattice parameters of $\beta_{12}$ - borophene are $a = |\vec{a}| = \sqrt{3}a_0$ and $b = |\vec{b}| = 3a_0$ where $a_0$ is the boron-boron atom distance. (c) Top view of the adsorption of the gas molecule on $\beta_{12}$ - borophene geometry. (d) 3D arrangement of the adsorption gas molecule on $a$ site of $\beta_{12}$ - borophene unit cell.
}
\label{fig1}
\end{figure}

The schematic illustration of $\beta_{12}$ - borophene field effect transistor-based gas sensor is shown in Fig. \ref{fig1}(a). The honeycomb lattice arrangement and the unit cell of $\beta_{12}$ - borophene is shown in Fig. \ref{fig1}(b). The unit cell comprises of five boron atoms labelled a, b, c, d and e. The $\beta_{12}$ - borophene lattice can be generated using the lattice basis vectors ($\vec{a}, \vec{b}$) $ = (\sqrt{3} a_0 \hat{e}_x, 3 a_0 \hat{e}_y)$ where $a_0 = 1.74\AA$ is the boron-boron atom distance \cite{xia}. It is to be noted that borophene does not allow the formation of in-plane $\sigma$ bonds due to the lack of one electron in the boron atom compared to the carbon atom. So, the C-atoms can be treated as a perfect donor in filling the in-plane hexagonal $\sigma$-bonds. Borophene has the energy levels involving $s$, $p_x$, $p_y$ and $p_z$ orbitals which are $sp^2$ - hybridized. However, only $p_z$ - orbitals can contribute to the carrier dynamics. It is due to the reason that the wave function at the Fermi energy has a vanishing amplitude at site c resulting in phase cancellation at the six-fold B-atoms. The band structure of $\beta_{12}$ - borophene in the presence of adsorbed gas molecules can be obtained using the tight binding approach considering only the nearest neighbour approximation. To model the molecular adsorption effect, we consider a $\beta_{12}$ - borophene sheet consisting of N sites. The gas adsorption in the $\beta_{12}$ - borophene lattice and the $n^\text{th}$ unit cell with its nearest neighbours is shown in Figs. \ref{fig1}(c) and \ref{fig1}(d). 

The matrix equation for the $n^\text{th}$ unit cell according to tight binding model can be written as \cite{dutta, pourasl}
\begin{equation}
\label{eq1}
\sum_m \left[\mathcal{H}_{nm}\right]\{\Psi_m\} = E\{\Psi_n\} 
\end{equation}
where $\{\Psi_n\}$ is a ($b \times 1$) column matrix corresponds to the wavefunction in the unit cell $n$. It is to be noted that the Hamiltonian matrix for pristine $\beta_{12}$ - borophene with five atoms in its unit cell will be a ($5 \times 5$) matrix. To characterize the effect of gas adsorption, we consider that each gas molecule is adsorbed only in each A site of the unit cell of $\beta_{12}$ - borophene as shown in Fig. \ref{fig1}(d). Thus, the Hamiltonian of the $\beta_{12}$ - borophene in the presence of a gas molecule will be a matrix of ($6 \times 6$) dimension.
 
We consider a plane waveform of the wave function, i.e., ${\Psi_n} = \Psi_0 e^{i \vec{k}.\vec{d}_n}$, with $\vec{k}$ and $\vec{d}_n$ being the plane wave vector and distance between a source with the $n^\text{th}$ unit-cell respectively. The band structure of the $\beta_{12}$ - borophene can be calculated by solving a matrix eigenvalue equation of the form \cite{dutta, pourasl}

\begin{equation}
\label{eq2}
[h(k)] = \sum_{nm} \left[\mathcal{H}_{nm}\right]e^{i\vec{k}.(\vec{d}_n - \vec{d}_m)}
\end{equation}
where $\{\Psi_n\}$ is a ($b \times 1$) column matrix corresponding to the wavefunction in the unit cell $n$. It is to be noted that the Hamiltonian matrix for pristine $\beta_{12}$ - borophene with five atoms in its unit cell will be a ($5 \times 5$) matrix. To characterize the effect of gas adsorption, we consider that each gas molecule is adsorbed only in each A site of the unit cell of $\beta_{12}$ - borophene as shown in Fig. \ref{fig1}(d). Thus, the Hamiltonian of the $\beta_{12}$ - borophene in the presence of a gas molecule will be a matrix of ($6 \times 6$) dimension.

\begin{widetext}
\begin{multline}
\label{eq3}
[h(k)]_{nm} = \left(
\begin{array}{cccccc}
 0 & t_{\text{ab}} \delta_k  & 0 & 0 & 0 & 0 \\
 0 & 0 & 0 & 0 & 0 & 0 \\
 0 &  t_{\text{cb}} \delta_k & 0 &  t_{\text{cd}} \delta_k & 0 & 0 \\
 0 & 0 & 0 & 0 & 0 & 0 \\
 0 & 0 & 0 &  t_{\text{ed}} \delta_k & 0 & 0 \\
 0 & 0 & 0 & 0 & 0 & 0 \\
\end{array}
\right)
+ \left(
\begin{array}{cccccc}
 0 & 0 & 0 & 0 & 0 & 0 \\
 0 & 0 & 0 & 0 & 0 & 0 \\
 0 & 0 & 0 & 0 & 0 & 0 \\
 0 & 0 & 0 & 0 & 0 & 0 \\
 t_{\text{ea}} & 0 & 0 & 0 & 0 & 0 \\
 0 & 0 & 0 & 0 & 0 & 0 \\
\end{array}
\right)
+ \left(
\begin{array}{cccccc}
 0 & 0 & 0 & 0 & 0 & 0 \\
  t_{\text{ba}} \delta_k^\ast & 0 &  t_{\text{bc}} \delta_k^\ast  & 0 & 0 & 0 \\
 0 & 0 & 0 & 0 & 0 & 0 \\
 0 & 0 &  t_{\text{dc}} \delta_k^\ast & 0 &  t_{\text{de}} \delta_k^\ast  & 0 \\
 0 & 0 & 0 & 0 & 0 & 0 \\
 0 & 0 & 0 & 0 & 0 & 0 \\
\end{array}
\right)
+
\left(
\begin{array}{cccccc}
 0 & 0 & 0 & 0 & t_{\text{ae}} & 0 \\
 0 & 0 & 0 & 0 & 0 & 0 \\
 0 & 0 & 0 & 0 & 0 & 0 \\
 0 & 0 & 0 & 0 & 0 & 0 \\
 0 & 0 & 0 & 0 & 0 & 0 \\
 0 & 0 & 0 & 0 & 0 & 0 \\
\end{array}
\right)
\end{multline}
\end{widetext}
where, $m = 1, 2, 3, 4$ correspond to the four nearest unit cells of the $n^\text{th}$ unit cell. Here, $t_\text{ij}$ are the hopping energy parameter of $i^\text{th}$ and $j^\text{th}$ boron atoms. The hopping parameters for the homogeneous model are $t_\text{ij} = -2$ eV \cite{huang}, while in the inversion non-symmetric (INS) model, the hopping parameters are: $t_{\text{ab}} = t_{\text{de}} = -2.04$ eV, $t_{\text{ac}} = t_{\text{ce}} = -1.79$ eV, $t_{\text{bc}} = t_{\text{cd}} = -1.84$ eV, $t_{\text{bd}} = -1.91$ eV, $t_{\text{ad}} = 0$ eV and $t_{\text{ae}} = -2.12$ eV \cite{huang}. Here, we define $\delta_k \equiv \exp\left[-\frac{i a_0 k_x}{2}\right]$. It is to be noted that one may use the lattice parameters ($\vec{a}, \vec{b}$) for the calculation of distance $\vec{d}_n$. However, in this work, the distance between the lattice points $(\vec{d}_m - \vec{d}_n)$ are obtained by converting the lattice parameters in terms of bond length $a_0$ for simplicity of the calculations.

In a similar way, the interaction of the $n^\text{th}$ unit cell with the adsorbed gas molecule can be written as 
\begin{equation}
\label{eq4}
[h(k)]_{nn}  = \left(
\begin{array}{cccccc}
 \varepsilon _\text{a} & t_{\text{ab}} & t_{\text{ac}} & 0 & 0 & t' \\
 t_{\text{ba}} &  \varepsilon _\text{b} & t_{\text{bc}} & t_{\text{bd}} & 0 & 0 \\
 t_{\text{ca}} & t_{\text{cb}} &  \varepsilon _\text{c} & t_{\text{cd}} & t_{\text{ce}} & 0 \\
 0 & t_{\text{db}} & t_{\text{dc}} &  \varepsilon _\text{d} & t_{\text{de}} & 0 \\
 0 & 0 & t_{\text{ec}} & t_{\text{ed}} &  \varepsilon _\text{e} & 0 \\
 t' & 0 & 0 & 0 & 0 &  \varepsilon'\\
\end{array}
\right)
\end{equation}
where, $\varepsilon_i$ are the corresponding on-site energy of the $i^\text{th}$ atom. The on-site energies in homogeneous model are $\varepsilon_i = 0$ eV while in the INS model on-site energies are: $\varepsilon_\text{a} = \varepsilon_\text{d} = 0.196$ eV, $\varepsilon_\text{b} = \varepsilon_\text{e} = -0.058$ eV and $\varepsilon_\text{c} = -0.845$ eV \cite{huang}. The parameters $t'$ and $\varepsilon'$ characterize the hopping energy between the Boron atom and the adsorbed molecule and the on-site energies, respectively.

\begin{figure}[hbt]
\centerline
\centerline{
\includegraphics[scale = 0.47]{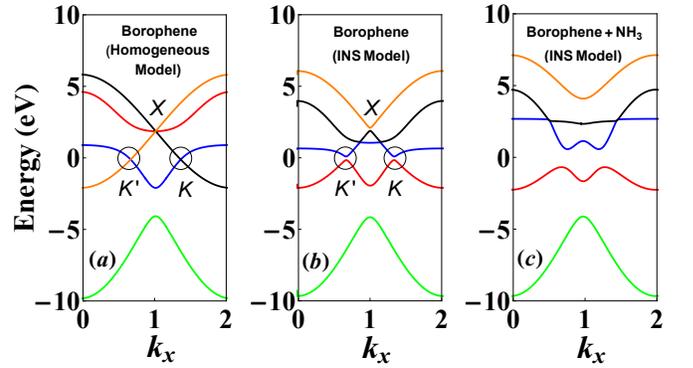}
\hspace{-0.09cm}
}
\caption{Band structure of $\beta_{12}$ - borophene using (a) homogenous and (b) the INS model. The plot (c) represents the band structure of $\beta_{12}$ - borophene in presence of NH$_3$ using the INS model.}
\label{fig2}
\end{figure}

Thus, the total tight binding Hamiltonian of the $\beta_{12}$-borophene system in presence of adsorbed gas in view of Eq. (\ref{eq3}) and Eq. (\ref{eq4}) can be expressed as
\begin{equation}
\label{eq5}
\left[h(k)\right] = \left(
\begin{array}{cccccc}
 \varepsilon _\text{a} & f^\ast_k  t_{\text{ab}} & t_{\text{ac}} & 0 & t_{\text{ae}} & t' \\
  f_k t_{\text{ab}} & \varepsilon _\text{b} & f_k t_{\text{bc}}  & t_{\text{bd}} & 0 & 0 \\
 t_{\text{ac}} & f^\ast_k t_{\text{bc}} & \varepsilon _\text{c} & f^\ast_k  t_{\text{cd}} & t_{\text{ce}} & 0 \\
 0 & t_{\text{bd}} & f_k t_{\text{cd}}  & \varepsilon _\text{d} & t_{\text{de}} f_k & 0 \\
 t_{\text{ae}} & 0 & t_{\text{ce}} & f^\ast_k  t_{\text{de}} & \varepsilon _\text{e} & 0 \\
 t' & 0 & 0 & 0 & 0 & \varepsilon'\\
\end{array}
\right)
\end{equation}
where, $f_k \equiv 1+\exp\left[\frac{i a_0 k_x}{2}\right]$.  The band structure and electronic dispersion of pristine $\beta_{12}$ - borophene can be obtained by diagonalizing the Hamiltonian. 

\begin{figure*}[hbt]
\centerline
\centerline{
\includegraphics[scale = 0.535]{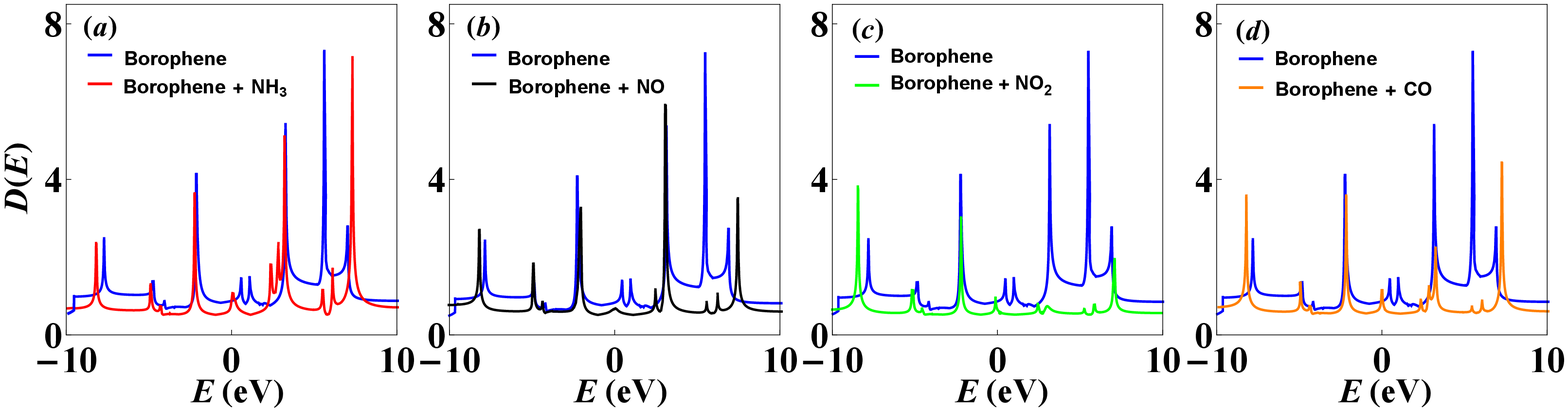}
\hspace{-0.09cm}
}
\caption{DOS for $\beta_{12}$ - borophene using INS model in absence and in presence of (a) NH$_3$, (b) NO (c) NO$_2$ and (d) CO  gases.}
\label{fig3}
\end{figure*}
For non-trivial solutions, we can write, 
\begin{equation}
\label{eq6}
det[h(k)-E_{k}\hat{I}]=0, \indent \lambda\in \lbrace 1,....5 \rbrace
\end{equation}
where, $E_{k}$ are the energy eigen values of the $\left[h(k)\right]$.  The determinant of the matrix can be expressed for pristine $\beta_{12}$ - borophene as 
\begin{equation}
\label{eq7}
E_k^5+\mathcal{Q}_4E_k^4 +\mathcal{Q}_3E_k^3 +\mathcal{Q}_2E_k^2 +\mathcal{Q}_1E_k+\mathcal{R} = 0
\end{equation}
\noindent where $\mathcal{Q}_1$, $\mathcal{Q}_2$, $\mathcal{Q}_3$, $\mathcal{Q}_4$ and $\mathcal{R}$ are the $k$ - dependent variable. An extensive form of the same is displayed in Appendix A. The energy band and the brillouin zones of $\beta_{12}$ - borophene can be obtained by solving the dispersion relation Eq. (\ref{eq7}). Fig. \ref{fig2}, presents the band structure of $\beta_{12}$ - borophene in both homogeneous and the INS model. It is evident from Fig. \ref{fig2} that $\beta_{12}$ - borophene has three conduction and two valence bands. So, $\beta_{12}$ - borophene has five bands. It is to be noted that the band edges touch at the high symmetry $K'$ and $K$ points with coordinates $(-2\pi/3a_0,0)$ and $(2\pi/3a_0,0)$ indicating the presence of Dirac fermions at $E_k = 0$. Also, triplet fermions at X and M points appear at $E_k \neq 0$. Moreover, a direct energy gap exists between $\sim 0.25$ eV at $K'$ and $K$ points for the INS model of $\beta_{12}$ - borophene. To understand the molecular gas adsorption effect on borophene energy, we have plotted the band structure of $\beta_{12}$ - borophene in the presence of NH$_3$ gas in Fig. \ref{fig1} (c). It is evident that the band gap at $K'$, $K$ and $X$ points increases in the presence of adsorbed NH$_3$ molecule. Moreover, it is to be noted that the widening of the band gap is dependent on the hopping energy $t'$ of the boron atom with the adsorbed gas molecule. 

The hopping energies can be calculated by using the relation \cite{pourasl}
\begin{eqnarray}
\label{eq8}
t_{xy} = t \left(\frac{a_0}{d_{xy}}\right)
\end{eqnarray}  
where, $t_{xy}$ and $d_{xy}$ are the hopping energy and the distance between $\beta_{12}$ - borophene surface and the gas molecule respectively. 

The density of states (DOS) has an energy dependence which signifies the number of allowed states per unit area at a given energy range $E$ and $E+dE$ and can be obtained by using the relation \cite{dutta}.
\begin{equation}
\label{eq9}
D(E) = \frac{\Delta n}{A\Delta E}
\end{equation}
\noindent where, A is the area of the $\beta_{12}$ - borophene surface and $n$ is the carrier concentration of the electrons. Although we obtain an analytic expression for the DOS using Eq. (\ref{eq7}) and Eq. (\ref{eq9}), in presence of adsorbed gas molecules it is too large to represent. 

The carrier concentration in the $\beta_{12}$ -  borophene based gas sensor can be obtained by using the standard relation \cite{dutta}
\begin{equation}
\label{eq10}
n(E) =  \int_0^\infty D(E) f(E) dE
\end{equation}
\noindent where, $f(E) = \frac{1}{1 + \exp\left(\frac{E-E_\text{F}}{k_\text{B} T}\right)}$ is the Fermi-Dirac distribution function, $E_\text{F}$ is the Fermi energy and $k_\text{B}$ is the Boltzmann constant. 

\begin{table}[]
\caption{Hopping parameter for gas molecules adsorbed on $\beta_{12}$ - Borophene surface}
\label{Tab1}
\begin{tabular}{|c|c|c|c|}
\hline
Adsorbed  & Adsorbate distance & Hopping energy  & On-site Energy\\
Molecule & from Borophene    & $t_{xy}$ (eV) & $\varepsilon'$ (eV) \\ 
& surface $d_{xy}$ (\AA)  & &  \\ \hline
NH$_3$               & 1.63        &      -2.113    &  1.11   \\ \hline
NO                & 1.38        &      -2.496  &  0.95   \\ \hline
NO$_2$               & 1.57        &      -2.194  &  1.75 \\ \hline
CO                & 1.48         &     -2.167    &  1.19     \\ \hline
\end{tabular}
\end{table}

\begin{figure*}[hbt]
\centerline
\centerline{
\includegraphics[scale = 0.54]{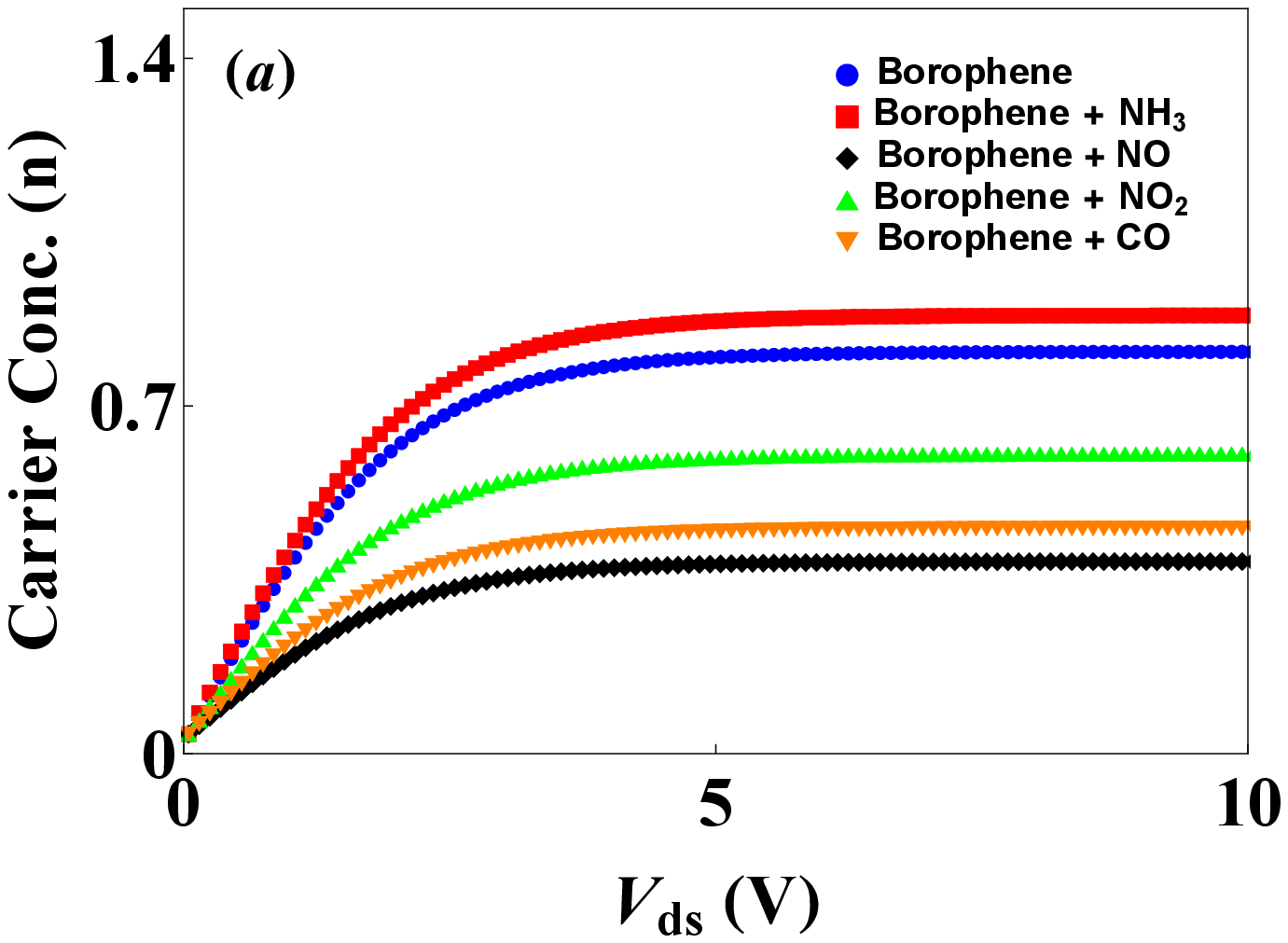}
\hspace{0.2cm}
\includegraphics[scale = 0.54]{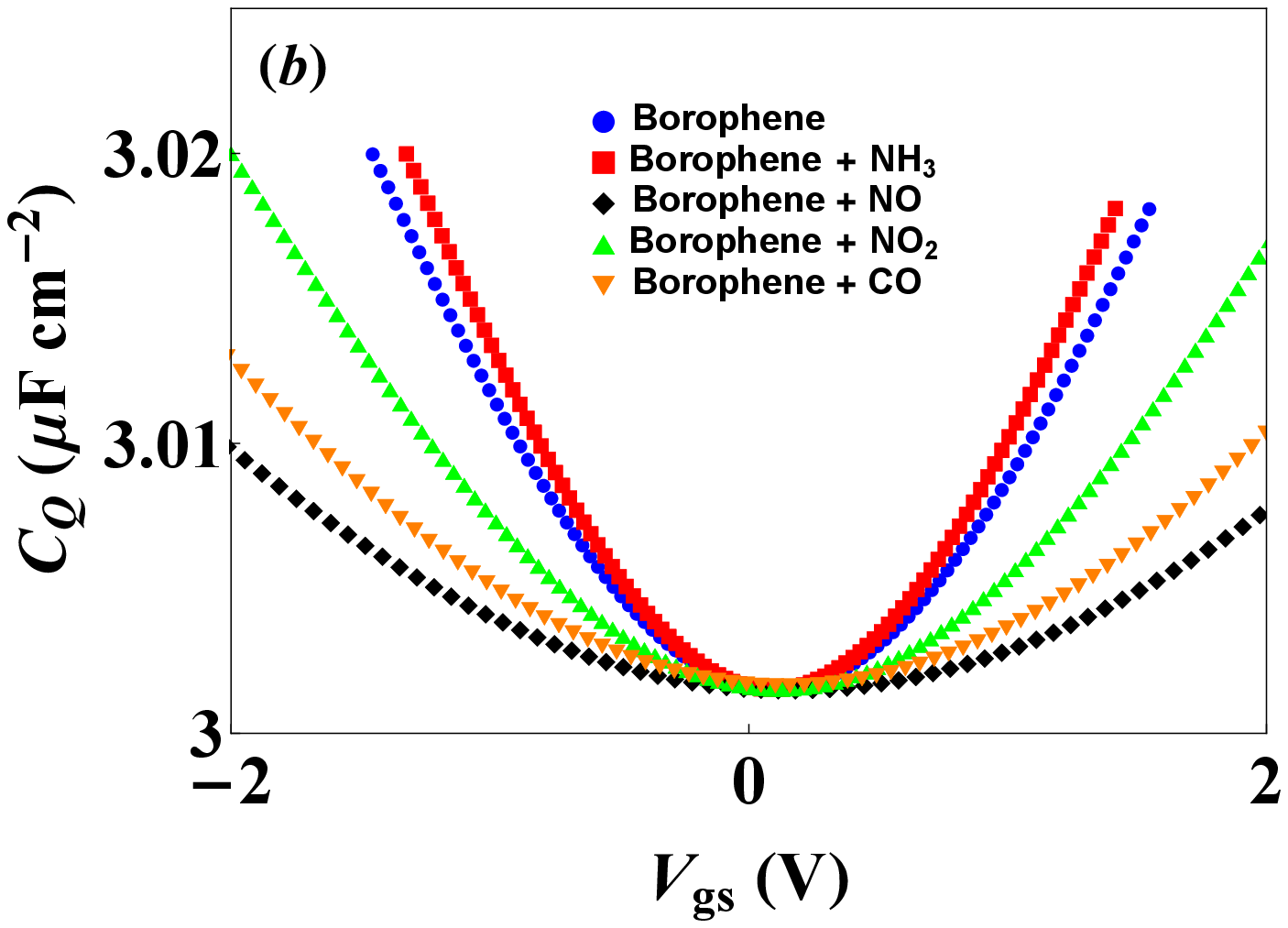}
}
\caption{(a) Variation of carrier concentration with $V_\text{ds}$ of $\beta_{12}$ - borophene in absence and in presence for different adsorbed gases. (b) Quantum conductance of $\beta_{12}$ - borophene as a function of $V_\text{gs}$ in absence and in presence of adsorbed gas molecules.}
\label{fig4}
\end{figure*}

\section{Results and Discussions}
\subsection{Density of states}
The density of states (DOS) of $\beta_{12}$ - borophene in presence of adsorbed gas molecules is studied using Eq. (\ref{eq10}) in Fig. \ref{fig3}. The hopping parameters  for different gases are calculated using Eq. (\ref{eq6}) and given in Tab. \ref{Tab1}.  We observe a significant change in the DOS spectra due to the adsorption of different gases. It is to be noted that the peak position in DOS spectra correspond to the flatter band energies in the $E-k$ diagram. Moreover, each extremum point can also be visualized as a Van-Hove singularity (VHS) in the DOS. At the same time, the smooth zones in the band structure depict the presence of localized electrons. As observed earlier from Fig. \ref{fig2}(c) that the presence of adsorbed NH$_{3}$ molecule drastically changes the band structure resulting in a significant change in DOS also. 
The change in the peak positions in DOS indicates the high sensitive gas sensing performance of $\beta_{12}$ - borophene. A similar characteristic is also observed for the DOS for pristine $\beta_{12}$ - borophene in the presence of adsorbed gases as depicted in Fig. \ref{fig3}. We consider the INS model for all our analysis, and the effect of the adsorbed gas molecule on DOS can be understood via the hopping energy parameter $t_{xy}$ from Eq. (\ref{eq8}). It is to be noted that the presence of more number of bands in the $E-k$ diagram indicates the presence of more degenerate bands in the DOS spectra. A similar characteristic in DOS is also for $\beta_{12}$ - borophene in the presence of NO molecule. However, in this case, the VHS at $E \sim 6.5$ eV shifted towards a higher energy region, as seen from Fig. \ref{fig3}(b). However, the DOS is found to be quite identical with $\beta_{12}$ - borophene for $E < 0$ regions. The adsorption of NO$_{2}$ molecule significantly changes the band structure, resulting in a significant change in the DOS. In this case, the VHS at $E \sim 3.5$ eV disappears completely, as seen in Fig. \ref{fig3}(c). The DOS for $\beta_{12}$ - borophene in the presence of CO molecule is similar to that for $\beta_{12}$ - borophene with adsorbed NO molecule. However, the peak positions shifted towards higher energy for adsorption of CO molecule, as seen from Figs. \ref{fig3}(d).

\subsection{Carrier Concentration and Quantum Capacitance}
To obtain an expression for quantum capacitance we consider the device is in quasi-equilibrium and the carrier distribution shifted by the local electrostatic potential. Then the charge density $Q$ of the electrons can be written as \cite{dutta}
\begin{equation}
\label{eq11}
Q = q\int_0^\infty D(E) f\left(E + E_\text{g} + q V_\text{a} \right)dE
\end{equation}
where, $q$ is the magnitude of the electronic charge and $V_\text{a}$ is the local electrostatic potential and $E_\text{g}$ is the band gap. The quantum capacitance $C_\text{Q}$ of the system is defined as \cite{hassanzadazar, yu3, li4} 
\begin{align}
\label{eq12}
C_\text{Q} &=  \frac{\partial Q}{\partial V_\text{a}} \nonumber\\
&= \frac{q}{4k_\text{B} T}\int_0^\infty D(E) \text{sech}\left(\frac{E + E_\text{g} + qV_\text{a} }{2 k_\text{B} T}\right) dE
\end{align}

In Fig. \ref{fig4}(a), we plot the quantum capacitance ($C_\text{Q}$) of $\beta_{12}$ - borophene FET against the gate-source voltage ($V_\text{gs}$) for various adsorbed gas molecules, considering their different hopping energy parameters. It is observed that at zero gate voltage the $C_\text{Q}$ is minimum while it increases gradually with increase or decrease in $V_\text{gs}$. The charge transfer from the adsorbed molecules results in the modification of the DOS of $\beta_{12}$ - borophene resulting in the change in the number of charge carrier concentration. This change in charge carrier in turn can change the $C_\text{Q}$ between the gate electrode and the $\beta_{12}$ - borophene surface. It is seen from the Fig. \ref{fig4}(b) that the $C_\text{Q}$ slightly increased when the adsorbed molecule was NH$_{3}$  while it decreased for NO, NO$_2$  and CO. This change in $C_\text{Q}$ will result in the variations in I-V characteristics of the $\beta_{12}$ - borophene.  

\subsection{I-V characteristics}
\begin{figure}[hbt]
\centerline
\centerline{
\includegraphics[scale = 0.5]{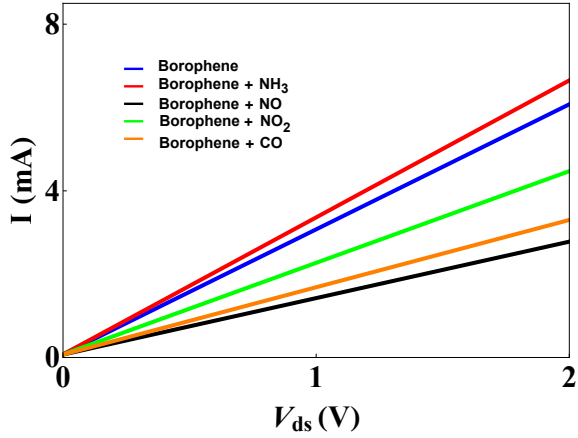}
\hspace{-0.09cm}
}
\caption{I-V characteristics of the $\beta_{12}$ - borophene FET in absence and in presence of different adsorbed gas molecules.}
\label{fig5}
\end{figure}

The I-V characteristics of the $\beta_{12}$ - borophene FET based sensor can be understood from quantum capacitance using the expression \cite{hassanzadazar, yu3, li4}
\begin{equation}
\label{eq13}
I = \mu C_\text{Q} V_\text{ds} E
\end{equation}
where, $\mu$ is the mobility of the electrons and $V_\text{ds}$ is the drain-source voltage. The FET based gas sensors respond to gas adsorption by adjusting the gate bias which controls the carrier concentration in the channel thus tuning the amount of charge carriers in exchange with the absorbed gas molecules. In order to confirm the influence of carrier concentration on the gas response, the response curves of FET-based sensors with $V_\text{ds}$ is shown in Fig \ref{fig5}. It has been clearly seen that the NO response on the surface, increases with the increase of $V_\text{ds}$.

The current – voltage (I-V) characteristics are calculated using our developed formalism, the results being shown in Fig \ref{fig5}. It reflects the effect of molecular adsorption of NH$_3$, NO, NO$_2$ and CO on the quantum capacitance of borophene. The sensitivities of the proposed FET-based sensor for NH$_3$, NO, NO$_2$ and CO increases with the increase in $V_\text{ds}$. Among the four gas molecules, the sensitivity of NH$_3$ is found to be highest for given $V_\text{ds}$, showing better sensing performance. The change in current has been observed after the adsorption of the gas molecules. The adsorption of the gas molecules on the $\beta_{12}$ - borophene surface could also change the energy bandgap of the $\beta_{12}$ - borophene which in turn would change the conductivity of the $\beta_{12}$ - borophene sensor. On the other hand, the modulation of the concentration of the charge carriers on the borophene surface occurs due to the charge transfer between borophene and gas molecules. The I-V characteristics are calculated for pristine $\beta_{12}$ - borophene and in presence of adsorbed gas molecules using Eq. (\ref{eq13}) and the results are shown in Fig. \ref{fig5}. It reflects the effect of molecular adsorption of NH$_{3}$, NO, NO$_{2}$, CO on the quantum capacitance of borophene. The sensitivities of the proposed FET-based sensor for NH$_{3}$, NO, NO$_{2}$, CO increase with the increase in $V_\text{ds}$. Among the four gas molecules, the sensitivity of NO is found to be highest for given $V_\text{ds}$, showing better sensing performance. The change in current has been observed after the adsorption of the gas molecules. The adsorption of the gas molecules on the borophene surface could also change the energy bandgap of the borophene which in turn would change the conductivity of the borophene sensor. On the other hand, the modulation of the concentration of the charge carriers on the borophene surface occurs due to the charge transfer between borophene and gas molecules.

\section{Conclusions}
In this work, we have studied the effects of molecular adsorption on the physical and electrical properties of $\beta_{12}$ - borophene. We propose a  theoretical model  based on tight binding technique to study the impact of various adsorbed gases on the number concentration, quantum capacitance and I-V characteristics of the FET based $\beta_{12}$ - borophene gas sensor. The study reveals the variation of borophene energy band gap near Fermi energy influenced by the adsorption of gas molecules. Furthermore, we study the density of states in the presence of different gases through the carrier concentration and developed the quantum capacitance model of FET based $\beta_{12}$ - borophene gas sensor.
The existence and shifting of the Van-Hove singularities at different energy sites in presence of gas molecules indicate the sensing performance of ability of $\beta_{12}$ - borophene. The study was mainly focused on the adsorption of NH$_{3}$, NO$_{2}$, NO and CO gas molecules and the effects on the  quantum capacitance and the I-V characteristics of the borophene sensor was  investigated. The present study specifies a significant variation in the band gap and quantum capacitance after gas adsorption contributing a change in the conductance, carrier concentration and current in the borophene gas sensor. This work paves the way for future studies of borophene as gas sensor due to its attractive gas sensing properties.

\appendix
\section{Extended form of $\mathcal{Q}$'s and $\mathcal{R}$}
In this section we present an extended form of  $\mathcal{Q}$'s and $\mathcal{R}$ appearing in Eq. (\ref{eq7}). For a pristine $\beta_{12}$ - borophene i.e., for $t' = 0$ and $\varepsilon' = 0$, the $\mathcal{Q}$'s and $\mathcal{R}$ are defined as 
\begin{widetext}
\begin{multline}
\mathcal{Q}_1 = \{16 t_{\text{ab}}^2 (t_{\text{cd}}^2+t_{\text{de}}^2)+16 t_{\text{bc}}^2 t_{\text{de}}^2\}\cos ^4\left(\frac{ka}{4}\right) +[8 t_{\text{ab}} t_{\text{ac}} \{t_{\text{bc}} (\varepsilon_\text{d}+\varepsilon_\text{e})-t_{\text{bd}} t_{\text{cd}}\}-4 \varepsilon_\text{a} \varepsilon_\text{b} t_{\text{cd}}^2-4 \varepsilon_\text{a} \varepsilon_\text{b} t_{\text{de}}^2+8 \varepsilon_\text{a} t_{\text{bc}} t_{\text{bd}} t_{\text{cd}}-4 t_{\text{bc}}^2 \varepsilon_\text{e} (\varepsilon_\text{a}
\\+\varepsilon_\text{d})
-4 \varepsilon_\text{a} t_{\text{bc}}^2 \varepsilon_\text{d}-4 \varepsilon_\text{a} \varepsilon_\text{c} t_{\text{de}}^2+8 \varepsilon_\text{a} t_{\text{cd}} t_{\text{ce}} t_{\text{de}}-4 \varepsilon_\text{a} t_{\text{cd}}^2 \varepsilon_\text{e}-8 t_{\text{ab}} t_{\text{ae}} (t_{\text{bc}} t_{\text{ce}}+t_{\text{bd}} t_{\text{de}})-4 t_{\text{ab}}^2 \varepsilon_\text{e} (\varepsilon_\text{c}+\varepsilon_\text{d})-4 t_{\text{ab}}^2 \varepsilon_\text{c} \varepsilon_\text{d}+4 t_{\text{ab}}^2 t_{\text{ce}}^2-8 t_{\text{ac}} t_{\text{ae}} t_{\text{cd}} t_{\text{de}}+4 t_{\text{ac}}^2 t_{\text{de}}^2
\\+4 t_{\text{ae}}^2 (t_{\text{bc}}^2+t_{\text{cd}}^2)-4 \varepsilon_\text{b} \varepsilon_\text{c} t_{\text{de}}^2+8 \varepsilon_\text{b} t_{\text{cd}} t_{\text{ce}} t_{\text{de}}-4 \varepsilon_\text{b} t_{\text{cd}}^2 \varepsilon_\text{e}+8 t_{\text{bc}} t_{\text{bd}} t_{\text{cd}} \varepsilon_\text{e}-8 t_{\text{bc}} t_{\text{bd}} t_{\text{ce}} t_{\text{de}}]\cos ^2\left(\frac{k a}{4}\right)-[\varepsilon_\text{d} \varepsilon_\text{e} \{\varepsilon_\text{c} (\varepsilon_\text{a}+\varepsilon_\text{b})+\varepsilon_\text{a} \varepsilon_\text{b}\}+\varepsilon_\text{a} \varepsilon_\text{b} \varepsilon_\text{c} \varepsilon_\text{d}
\\+\varepsilon_\text{a} \varepsilon_\text{b} \varepsilon_\text{c} \varepsilon_\text{e}+\varepsilon_\text{a} \varepsilon_\text{b} t_{\text{ce}}^2-t_{\text{bd}}^2 \varepsilon_\text{e} (\varepsilon_\text{a}+\varepsilon_\text{c})-\varepsilon_\text{a} t_{\text{bd}}^2 \varepsilon_\text{c}-\varepsilon_\text{a} t_{\text{ce}}^2 \varepsilon_\text{d}+2 t_{\text{ac}} t_{\text{ae}} t_{\text{ce}} (\varepsilon_\text{b}+\varepsilon_\text{d})-t_{\text{ac}}^2 t_{\text{ac}}^2 t_{\text{bd}}^2 \varepsilon_\text{e} -(\varepsilon_\text{b}+\varepsilon_\text{d})-t_{\text{ac}}^2 \varepsilon_\text{b} \varepsilon_\text{d}-t_{\text{ae}}^2 \varepsilon_\text{d} (\varepsilon_\text{b}+\varepsilon_\text{c})-t_{\text{ae}}^2 \varepsilon_\text{b} \varepsilon_\text{c}
\\+t_{\text{ae}}^2 t_{\text{bd}}^2-\varepsilon_\text{b} t_{\text{ce}}^2 \varepsilon_\text{d}+t_{\text{bd}}^2 t_{\text{ce}}^2]
\end{multline}
\begin{multline}
\mathcal{Q}_2 = \{4 t_{\text{bc}}^2 (\varepsilon_\text{a}+\varepsilon_\text{d}+\varepsilon_\text{e})+4 \varepsilon_\text{a} t_{\text{cd}}^2+4 \varepsilon_\text{a} t_{\text{de}}^2-8 t_{\text{ab}} t_{\text{ac}} t_{\text{bc}}+4 t_{\text{ab}}^2 (\varepsilon_\text{c}+\varepsilon_\text{d}+\varepsilon_\text{e})+4 \varepsilon_\text{b} t_{\text{cd}}^2+4 \varepsilon_\text{b} t_{\text{de}}^2-8 t_{\text{bc}} t_{\text{bd}} t_{\text{cd}}+4 \varepsilon_\text{c} t_{\text{de}}^2-8 t_{\text{cd}} t_{\text{ce}} t_{\text{de}}
\\+4 t_{\text{cd}}^2 \varepsilon_\text{e}\}\cos ^2\left(\frac{k a}{4}\right)-\varepsilon_\text{a} \varepsilon_\text{e} (\varepsilon_\text{b}+\varepsilon_\text{c}+\varepsilon_\text{d})-\varepsilon_\text{a} \varepsilon_\text{b} \varepsilon_\text{c}-\varepsilon_\text{a} \varepsilon_\text{b} \varepsilon_\text{d}+t_{\text{bd}}^2 (\varepsilon_\text{a}+\varepsilon_\text{c}+\varepsilon_\text{e})-\varepsilon_\text{a} \varepsilon_\text{c} \varepsilon_\text{d}+\varepsilon_\text{a} t_{\text{ce}}^2-2 t_{\text{ac}} t_{\text{ae}} t_{\text{ce}}+t_{\text{ac}}^2 (\varepsilon_\text{b}+\varepsilon_\text{d}+\varepsilon_\text{e})+t_{\text{ae}}^2 (\varepsilon_\text{b}
\\+\varepsilon_\text{c}+\varepsilon_\text{d})-\varepsilon_\text{d} \varepsilon_\text{e} (\varepsilon_\text{b}+\varepsilon_\text{c})-\varepsilon_\text{b} \varepsilon_\text{c} \varepsilon_\text{d}-\varepsilon_\text{b} \varepsilon_\text{c} \varepsilon_\text{e}+\varepsilon_\text{b} t_{\text{ce}}^2+t_{\text{ce}}^2 \varepsilon_\text{d}
\end{multline}
\begin{multline}
\mathcal{Q}_3 = -4 \cos ^2\left(\frac{k a}{4}\right) \left(t_{\text{ab}}^2+t_{\text{bc}}^2+t_{\text{cd}}^2+t_{\text{de}}^2\right)-t_{\text{ac}}^2-t_{\text{ae}}^2-t_{\text{bd}}^2-t_{\text{ce}}^2+\varepsilon_\text{a} \varepsilon_\text{b}+\varepsilon_\text{a} \varepsilon_\text{c}+\varepsilon_\text{a} \varepsilon_\text{d}+\varepsilon_\text{a} \varepsilon_\text{e}+\varepsilon_\text{b} \varepsilon_\text{c}+\varepsilon_\text{b} \varepsilon_\text{d}+\varepsilon_\text{b} \varepsilon_\text{e}+\varepsilon_\text{c} \varepsilon_\text{d}+\varepsilon_\text{c} \varepsilon_\text{e}
\\+\varepsilon_\text{d} \varepsilon_\text{e}
\end{multline}
\begin{equation}
\mathcal{Q}_4 = -\left(\varepsilon_\text{a}+\varepsilon_\text{b}+\varepsilon_\text{c}+\varepsilon_\text{d}+\varepsilon_\text{e}\right)
\end{equation}
\begin{multline}
\mathcal{R} = \cos ^2\left(\frac{ka}{4}\right)\{4 \varepsilon_\text{a} \varepsilon_\text{b} \varepsilon_\text{c} t_{\text{de}}^2-8 \varepsilon_\text{a} \varepsilon_\text{b} t_{\text{cd}} t_{\text{ce}} t_{\text{de}}+4 \varepsilon_\text{a} \varepsilon_\text{b} t_{\text{cd}}^2 \varepsilon_\text{e}-8 \varepsilon_\text{a} t_{\text{bc}} t_{\text{bd}} t_{\text{cd}} \varepsilon_\text{e}+8 \varepsilon_\text{a} t_{\text{bc}} t_{\text{bd}} t_{\text{ce}} t_{\text{de}}+4 \varepsilon_\text{a} t_{\text{bc}}^2 \varepsilon_\text{d} \varepsilon_\text{e}-8 t_{\text{ab}} t_{\text{ac}} t_{\text{bc}} \varepsilon_\text{d} \varepsilon_\text{e}
\\+8 t_{\text{ab}} t_{\text{ac}} t_{\text{bd}} (t_{\text{cd}} \varepsilon_\text{e}-t_{\text{ce}} t_{\text{de}})+8 t_{\text{ab}} t_{\text{ae}} t_{\text{bc}} t_{\text{ce}} \varepsilon_\text{d}+8 t_{\text{ab}} t_{\text{ae}} t_{\text{bd}} (\varepsilon_\text{c} t_{\text{de}}-t_{\text{cd}} t_{\text{ce}})-4 t_{\text{ab}}^2 \varepsilon_\text{d} (t_{\text{ce}}^2-\varepsilon_\text{c} \varepsilon_\text{e})+8 t_{\text{ac}} t_{\text{ae}} \varepsilon_\text{b} t_{\text{cd}} t_{\text{de}}-8 t_{\text{ac}} t_{\text{ae}} t_{\text{bc}} t_{\text{bd}} t_{\text{de}}-4 t_{\text{ac}}^2 \varepsilon_\text{b} t_{\text{de}}^2
\\-4 t_{\text{ae}}^2 \varepsilon_\text{b} t_{\text{cd}}^2+8 t_{\text{ae}}^2 t_{\text{bc}} t_{\text{bd}} t_{\text{cd}}-4 t_{\text{ae}}^2 t_{\text{bc}}^2 \varepsilon_\text{d}\}+\cos ^2\left(\frac{k a}{4}\right)^2-16 \varepsilon_\text{a} t_{\text{bc}}^2 t_{\text{de}}^2+32 t_{\text{ab}} t_{\text{ac}} t_{\text{bc}} t_{\text{de}}^2-32 t_{\text{ab}} t_{\text{ae}} t_{\text{bc}} t_{\text{cd}} t_{\text{de}}-16 t_{\text{ab}}^2 \varepsilon_\text{c} t_{\text{de}}^2+32 t_{\text{ab}}^2 t_{\text{cd}} t_{\text{ce}} t_{\text{de}}
\\-16 t_{\text{ab}}^2 t_{\text{cd}}^2 \varepsilon_\text{e})+\varepsilon_\text{a} \varepsilon_\text{c} \varepsilon_\text{e} (t_{\text{bd}}^2-\varepsilon_\text{b} \varepsilon_\text{d})+\varepsilon_\text{a} \varepsilon_\text{b} t_{\text{ce}}^2 \varepsilon_\text{d}-\varepsilon_\text{a} t_{\text{bd}}^2 t_{\text{ce}}^2-2 t_{\text{ac}} t_{\text{ae}} \varepsilon_\text{b} t_{\text{ce}} \varepsilon_\text{d}+2 t_{\text{ac}} t_{\text{ae}} t_{\text{bd}}^2 t_{\text{ce}}+t_{\text{ac}}^2 \varepsilon_\text{e} (\varepsilon_\text{b} \varepsilon_\text{d}-t_{\text{bd}}^2)+t_{\text{ae}}^2 \varepsilon_\text{c} (\varepsilon_\text{b} \varepsilon_\text{d}-t_{\text{bd}}^2).
\end{multline}
\end{widetext}


\begin{thebibliography}{99}
\bibitem{novoselove}
K. S. Novoselov, et. al., Nature {\bf 490}, 192-200 (2012).
\bibitem{lemme}
M. C. Lemme, et. al., Nat. Commun. {\bf 13}, 1392 (2022).
\bibitem{liu}
C. Liu, et. al.,  Nat. Nanotechnol. {\bf 15}, 545–557 (2020).
\bibitem{xu}
M. Xu, et. al., Chem. Rev. {\bf 113}, 3766 (2013).
\bibitem{quellmalz}
A. Quellmalz, et. al., Nat. Commun. {\bf 12}, 917 (2021).
\bibitem{illarionov}
Y. Y. U. Illarionov, et. al., Nat. Commun. {\bf 11}, 3385 (2020).
\bibitem{passi}
V. Passi, et. al., Adv. Mater. Interfaces {\bf 6}, 1801285 (2019).
\bibitem{fuechsle}
M. Fuechsle, et. al., Nat. Nanotechnol. {\bf 7}, 242–246 (2012).
\bibitem{acharjee}
S. Acharjee, et. al., Chaos {\bf 33}, 013136 (2023).
\bibitem{balandin}
A. A. Balandin, et. al., Nano Lett. {\bf 8}(3), 902-907 (2008).
\bibitem{lee}
C. Lee, et. al., Science {\bf 321}(5887), 385-388 (2008).
\bibitem{zhang}
N. Zhang, et. al., 2D Materials {\bf 5}(4), 045004 (2018). 
\bibitem{sang}
M. Sang, et. al., Nanomaterials (Basel). {\bf 9}(3), 374. (2019). 
\bibitem{pop}
E. Pop, V.  Varshney, and A. Roy, MRS Bulletin, {\bf 37}(12), 1273-1281 (2012).
\bibitem{papageorgiou}
D. G. Papageorgiou, I. A. Kinloch and R. J. Young, Prog. in Mat. Sci. {\bf 90}, 75-127 (2017).
\bibitem{neto}
A. H. C. Neto, Rev. Mod. Phys. {\bf 81}, 109 (2009).
\bibitem{mccann}
E. McCann and M. Koshino, Rep. Prog. Phys. {\bf 76} 056503 (2013).
\bibitem{ando}
T. Ando, NPG Asia Mater {\bf 1}, 17–21 (2009).
\bibitem{akhtar}
M. Akhtar, et. al., npj 2D Mater. Appl. {\bf 1}, 5 (2017).
\bibitem{vogt}
P. Vogt, et. al., Phys. Rev. Lett. {\bf 108}(15), 155501 (2012).
\bibitem{zhu2}
F. F. Zhu, et. al., Nat. Mater. {\bf 14}(10), 1020 (2015).
\bibitem{ji}
J. Ji, et. al., Nat. Commun. {\bf 7}, 13352 (2016).
\bibitem{drozdov}
I. K. Drozdov, et. al., Nat. Phys. {\bf 10}, 664 (2014).
\bibitem{davila}
M. Davila, et. al., New. J. Phys. {\bf 16}(9), 095002 (2014).
\bibitem{splendiani}
A. Splendiani, et. al., Nano. Lett. {\bf 10}, 1271 (2010).
\bibitem{mannix}
A. J. Mannix, et. al., Science {\bf 350}, 1513-1516 (2015).
\bibitem{zhang2}
Z. Zhang, et. al., Sci. Adv. {\bf 5}, eaax0246  (2019).
\bibitem{feng}
B. Feng, et. al., Phys. Rev. B {\bf 94}, 041408 (2016).
\bibitem{zhang3}
Z. Zhang, et. al., Adv. Funct. Mater. {\bf 27}, 1605059 (2017).
\bibitem{mozvashi}
S. M. Mozvashi, et al., Sci. Rep. {\bf 11}, 7547 (2021).
\bibitem{yuan}
j. Yua, et. al., RSC Adv., {\bf 7}, 8654-8660 (2017).
\bibitem{wang}
X. Wang et. al., Mater. Res. Express {\bf 8}, 065003 (2021).
\bibitem{shukla}
V. Shukla, et. al., J. Phys. Chem. C {\bf 121}, 26869–26876 (2017).
\bibitem{huang}
C. S. Huang, et. al., J. Phys. Chem. C {\bf 122}(26), 14665-14670 (2018).
\bibitem{hou}
C. Hou, C., et al. Nano Res. {\bf 15}, 2537–2544 (2022).
\bibitem{shen}
J. Shen, et. al., Appl. Sur. Sci. {\bf 504}, 144412 (2020).
\bibitem{peng}
B. Peng, et. al., Mater. Res. Lett. {\bf 5}(6), 399 (2017).
\bibitem{li2}
P. T. T. Le, T. C. Phong and M. Yarmohammadi, Phys. Chem. Chem. Phys. {\bf 21}, 21790 (2019).
\bibitem{nobahari}
M. M. Nobahari, RSC Adv. {\bf 12}, 648 (2022).
\bibitem{opoku}
F. Opoku, P. P. Govender, J. Phys. Chem. A {\bf 124}(11), 2288-2300 (2020).
\bibitem{tsu}
R. Tsu, Superlattice to Nanoelectronics (Second Edition), (2011).
\bibitem{xia}
J. Xia, et. al., Nat. Nanotechnol. {\bf 4}, 177 (2019).
\bibitem{hassanzadazar}
M. Hassanzadazar and K. Hadidi, Mater. Res. Express  {\bf 7}, 065006 (2020).
\bibitem{yu3}
G. L. Yu, et. al., Proc. Natl. Acad. Sci. {\bf 110}, 3282-3286 (2013).
\bibitem{li4}
J. K. Li,  et. al., Appl. Phys. Lett. 122, 063102 (2023).
\bibitem{inhofer}
A. Inhofer, Phys. Rev. Applied {\bf 9}, 024022 (2018).
\bibitem{cheremisin}
M. V. Cheremisin, Physica E: Low. Dimens. Syst. Nano. Struct. {\bf 69}, 153-158 (2015).
\bibitem{dutta}
S. Dutta, Quantum transport: Atom to Transistor, Cambridge University Press. U. K. (2005).
\bibitem{pourasl}
A. H. Pourasl, et. al. IEEE Sensors {\bf 19}, 10 (2019).
\bibitem{john}
D. L. John, L. C. Castro, and D. L. Pulfrey, J. Appl. Phys. {\bf 96}, 5180 (2004).
\bibitem{kolavada}
H. Kolavada, et. al., Electrochim. Acta {\bf 439}, 141589 (2023).
\end{thebibliography}
\end{document}